\documentclass[useAMs,a4paper]{mn2e}
\usepackage{savesym}
\usepackage{graphicx}
\expandafter\let\csname equation*\endcsname\relax
  \expandafter\let\csname endequation*\endcsname\relax 
\usepackage{subfig}
\usepackage{amsmath}
\usepackage{amssymb}
\usepackage{verbatim}
\usepackage[yyyymmdd,hhmmss]{datetime}
\usepackage{array}
\usepackage{times}
\usepackage[total={17.8cm,24.0cm},centering]{geometry} 
\usepackage{color}

\newcommand{\be}{\begin{equation}}
\newcommand{\beq}{\begin{equation}}
\newcommand{\ee}{\end{equation}}
\newcommand{\eeq}{\end{equation}}
\newcommand{\eea}{\end{eqnarray}}
\newcommand{\bea}{\begin{eqnarray}}

\title[A TDE about a maximally rotating $10^9 M_{\odot}$ BH]{ASASSN-15lh: a TDE about a maximally rotating $10^9 M_{\odot}$ black hole}
\author [Andrew Mummery, Steven A. Balbus]{Andrew Mummery\thanks{E-mail:
andrew.mummery@physics.ox.ac.uk}, {Steven A. Balbus}
\\
Oxford Astrophysics, Denys Wilkinson Building, Keble Road, Oxford, OX1 3RH, United Kingdom}
\begin{document}

\date{}

\pagerange{\pageref{firstpage}--\pageref{lastpage}} \pubyear{2020}

\maketitle

\label{firstpage}

\begin{abstract} 
We model the light curves of the novel and extremely luminous transient ASASSN-15lh at nine different frequencies, from infrared to ultra-violet photon energies, as an evolving relativistic disc produced in the aftermath of a  tidal disruption event.  {Good fits to all nine light curves are simultaneously obtained} when $M_{\rm acc} \simeq 0.07 M_{\odot}$ is accreted onto a black hole of mass $M \simeq 10^9 M_{\odot}$,  and near maximal rotation  $a/r_g = 0.99$.  The best fit black hole mass is consistent with a number of existing estimates from galactic scaling relationships.  {If confirmed,} our results represent the detection of one of the most massive rapidly spinning black holes to date, and are strong evidence for a TDE origin for ASASSN-15lh.   This would be the first TDE to be observed in the disc dominated state at optical and infrared frequencies. 
\end{abstract}

\begin{keywords}
accretion, accretion discs --- black hole physics --- tidal disruption events
\end{keywords}
\noindent
%Complied at \today\ \currenttime\ .

\section{Introduction}
\begin{figure}
  \includegraphics[width=.5\textwidth]{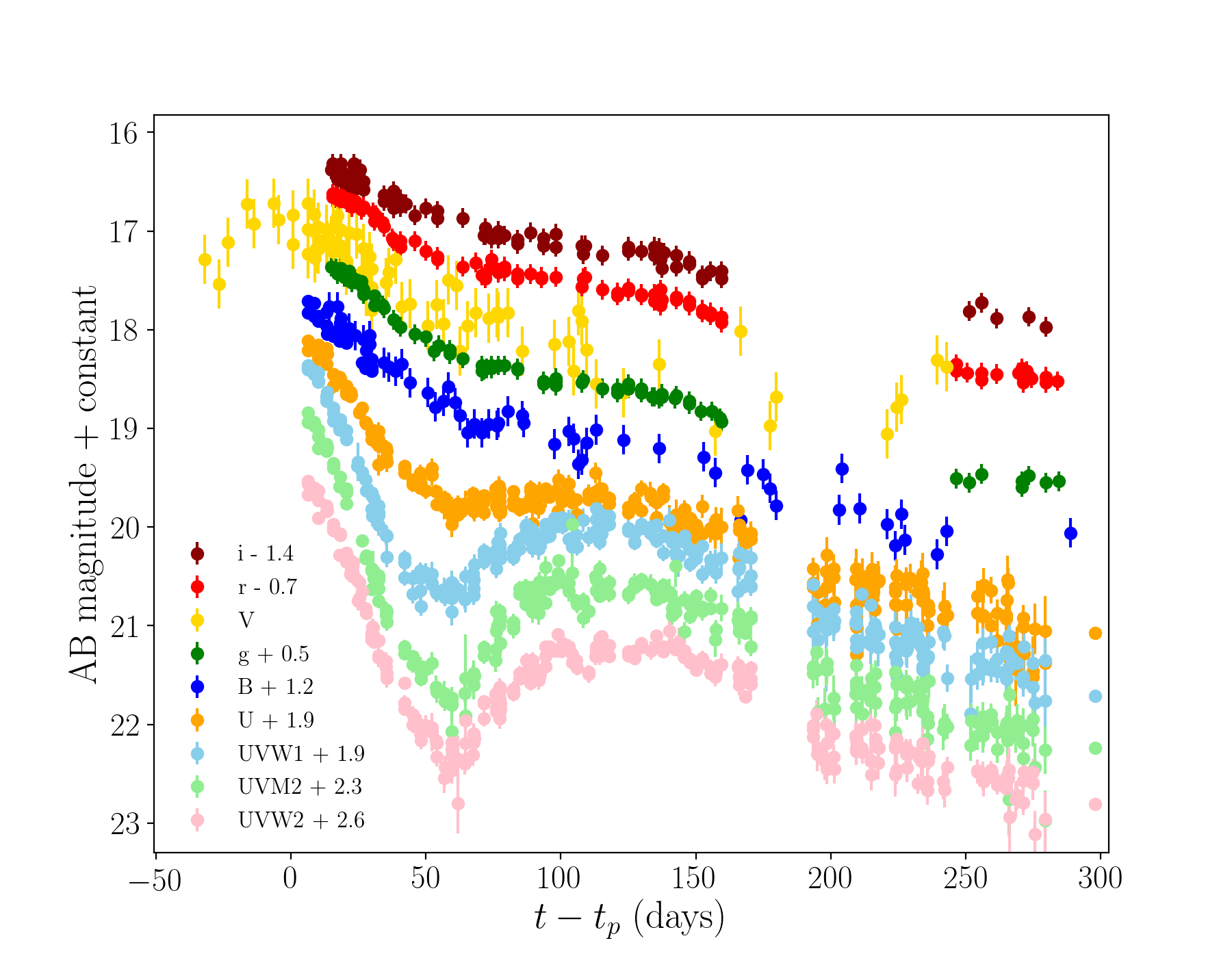} 
 \caption{The observed light curves of ASASSN-15lh. }
 \label{data}
\end{figure}

ASASSN-15lh is an extremely luminous transient event with red-shift $z = 0.2326$ (Dong {\it et al}. 2016) whose physical origin is contested.   Initially classified as a super-luminous supernova (Dong {\it et al}. 2016), it has since been argued that the properties of the host galaxy and the evolving spectrum of ASASSN-15lh are more consistent with a tidal disruption event (TDE) (Leloudas {\it et al}. 2016; hereafter L16).  The first 350 days of observations of ASASSN-15lh are shown in figure \ref{data}.  ASASSN-15lh is notable for its extreme brightness ($M_V = -23.5$ mag) and its double-humped UV light curves. It has been suggested that the properties of these UV light curves are supportive of a TDE origin for ASASSN-15lh, with the re-brightening corresponding to a transition to an accretion disc dominated state (L16).  

The current authors have studied the solutions of the thin disc evolution equation for Kerr spacetime geometries (Mummery \& Balbus 2019a,b), and have developed a model for the evolution of TDE light curves in their disc-dominated phase (Mummery \& Balbus 2020, hereafter MB20).  It is the purpose of this {\em Letter} to investigate whether a disc model can reproduce the properties of the evolving ASASSN-15lh light curves.   This will both serve as a test of the disc model in more novel accretion regimes, and as a test of the ASASSN-15lh TDE paradigm.   We model all nine ASASSN-15lh light curves displayed in figure \ref{data}, corresponding to observed photon energies ranging from the infrared to ultra-violet energies.  It is demonstrated that all 9 light curves transition to a disc-dominated state at $t \simeq t_{\rm peak} + 65$ days. These light curves are all well-described by an evolving disc model for the next $\sim 300$ days (figs \ref{spectrum}, \ref{fit}).  Interestingly, the best fit black hole parameters are those of a maximally rotating black hole of mass $M \simeq 10^9 M_{\odot}$.  This black hole mass is consistent with four independent estimates from galactic scaling relationships (L16, Kr{\" u}hler {\it et al.} 2018, Wevers {\it et al.} 2018).  For large black hole masses, rapid spins are strongly favoured, as only this regime allows for the tidal forces required to disrupt solar type stars outside of the black holes event horizon.  Our results are strongly supportive of a TDE origin for ASASSN-15lh, which would thus become the first TDE to be observed in the disc dominated state at optical and infrared frequencies.

\section{Light curve of a massive black hole}\label{scaling}

The host galaxy of ASASSN-15lh is a massive red galaxy with a small rate of ongoing star formation (L16).   Measurements of the galactic bulge mass, bulge luminosity and velocity dispersion (L16, Kr{\" u}hler {\it et al.} 2018, Wevers {\it et al.} 2018) suggest that the central black hole in ASASSN-15lh must be extremely massive, $M \sim 10^9 M_{\odot}$.   We show below that a detailed analysis of the ASASSN-15lh light curves reveals that they are indeed best fit by a $M \simeq 10^9 M_{\odot}$ black hole.    For the present, however, we note that the black hole properties may be inferred rather simply from qualitative features of the observed UV and optical light curves alone.  

The light curves of TDEs at ultraviolet energies show a transition to a disc dominated state at times $t \sim 100$--$300$ days after the initial disruption (van Velzen {\it et al}. 2019, MB20).  These same studies have likewise demonstrated that in this disc-dominated phase the light curves at ultra-violet energies are extremely flat with time, well approximated by $F_{UV} \sim {\rm const}$.  ASASSN-15lh shows a different, strongly time varying behaviour after the light curves transition at $t - t_{\rm peak} \sim 65$ days.  

This at first seems inconsistent with the results of van Velzen {\it et al}. (2019) and  MB20. 
 However, the ``standard'' flat UV light curves follows from a particular condition ordinarily satisfied by TDE discs (MB20):
\beq
E_{UV} \ll k_B f_\gamma T_{\rm peak}. 
\eeq
Here, $E_{UV}$ is the energy of the UV photons, and $f_\gamma T_{\rm peak}$ is the product of the maximum disc temperature $T_{\rm peak}$ and the red-shift factor $f_\gamma$ of the disc photons emitted in that region.  {The red-shift factor relates the observed energy of a photon to its emitted energy $f_\gamma \equiv E_{\rm obs}/E_{\rm emit}$, and is given by equation (8) of MB20. }

If, on the other hand,
\beq
E \gtrsim k_B f_\gamma T_{\rm peak}
\eeq
the observed light curves at these wavelengths evolve more rapidly than the power-law fall off associated with the disc's bolometric luminosity (MB20).     If ASASSN-15lh  truly is a disc-dominated TDE, then rapidly evolving UV light curves imply the condition:
\beq\label{condition}
f_\gamma T_{\rm peak} \lesssim E_{UV}/k_B \sim   5 \times 10^4 \, {\rm K} .
\eeq
X-ray observations suggest that standard TDE peak disc temperatures lie at $T_{\rm peak} \sim 10^6$ K (e.g., fig 4. MB20).   A drop in peak temperatures of a factor $\gtrsim 20$ suggests atypical, rather extreme parameters associated with the disc. 

In determining the magnitude of the physical peak disc temperature $T_{\rm peak}$,  there are to leading order 3 degrees of freedom within a disc model: i) the disc mass $M_d$; ii) the turbulent stress (an `$\alpha$' type parameter);  and iii) the black hole mass $M$.  {While one expects a spread of disc masses and $\alpha$ parameters, it is unlikely that either will vary by the orders of magnitude necessary to produce the ASASSN-15lh peak temperature change.  SMBH masses, however, certainly span many orders of magnitude, and it is to the black hole parameters that we naturally turn for an explanation of the low peak disc temperature. }

If the luminosity is a fixed fraction of the Eddington value, the disc temperature has a modest dependence on black hole mass $T \propto M^{-1/4}$.    X-ray bright TDEs (those with the hottest observed disc temperatures) are typically associated with black holes masses of $M \sim 10^6$--$10^7 M_{\odot}$  (Wevers {\it et al}. 2018).  To reduce the peak disc temperatures to UV energies, ASASSN-15lh must contain an unusually massive black hole.  
\begin{figure}
  \includegraphics[width=.49\textwidth]{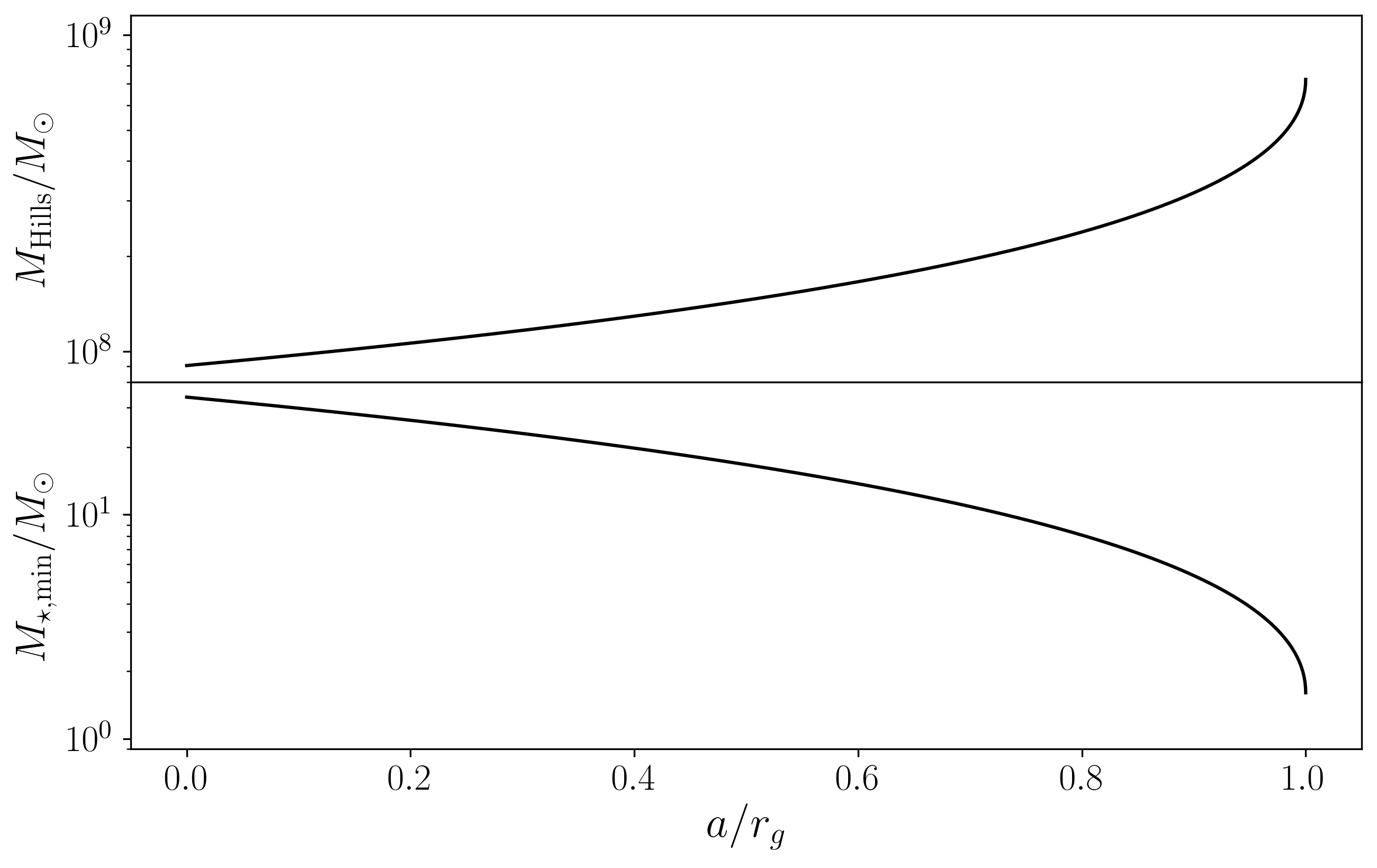} 
 \caption{Upper: The Hills mass, defined as the maximum black hole mass which can disrupt a solar  ($M_\star = M_{\odot}, R_\star = R_\odot$) star, as a function of black hole spin $a$. Lower: the minimum stellar mass of a solar type star ($R_\star = R_\odot (M_\star/M_\odot)^{\xi}, \xi = 0.79$)  which can be tidally disrupted by a $M = 10^9 M_{\odot}$ black hole as a function of black hole spin.   }
 \label{spin}
\end{figure}

While this argument is consistent with the large black hole mass inferred from host galaxy observations, it also implies a second condition: the black hole must be near maximally rotating. This is because the inferred black hole mass of ASASSN-15lh lies above the Hills mass (Dong {\it et al.} 2016, L16), defined as the black hole mass at which a Schwarzschild black hole would first directly capture a solar type star without tidally disrupting it:  $M_{\rm Hills} \simeq 9 \times 10^7 M_{\odot}$ (Hills 1975).  The Hills mass increases by roughly an order of magnitude for extreme Kerr spins (Kesden 2012), since a maximally rotating Kerr black hole has both a stronger tidal potential as well as a smaller event horizon.   Fig \ref{spin} shows the analytical model of Kesden (2012),  which predicts that a maximally rotating $10^9 M_{\odot}$ black hole can disrupt a $M_\star \gtrsim 1.5 M_{\odot}$ solar type star.   This is probably a conservative estimate, as it neglects stellar rotation and the bulge distortion of the star's surface on its approach.   Indeed, both hydrodynamical (Ivanov {\it et al}. 2006) and SPH (Sponholz 1994) simulations show the disruption of a $1 M_{\odot}$ by a maximally rotating $10^9 M_{\odot}$ black hole.    The lower half of figure \ref{spin} shows the minimum mass of a solar-type star (assumed to satisfy $R_\star \propto M_\star ^{0.79}$, Kippenhahn \& Weigert 1990), as a function of black hole spin, that can be tidally disrupted by a $10^9 M_{\odot}$ black hole.    

{The tidal disruption cross section increases slightly for more massive stars. A simple estimate (Magorrian \& Tremaine 1999) implies $\sigma_{\rm TDE} \propto R_\star M_\star^{-1/3} \propto M_\star^{0.46}$.  The enlarged cross section is countered  by the sparcity of high mass stars (${\rm d}N_\star/{\rm d} M_\star \propto M_\star^{-2.35}$ for a Salpeter mass function).   Thus, TDEs involving large stellar masses are expected to be rare ${\rm d} N_{\rm TDE}/{\rm d}M_\star \propto M_\star^{-1.89}$ (Stone \& Metzger 2016), particularly for early type galaxies like the host of ASASSN-15lh (L16).  This suggests that ASASSN-15lh contains a massive black hole with a large prograde spin.  } 

While large prograde black hole spins increase the physical peak disc temperature $T_{\rm peak}$, the innermost regions of a highly spinning black hole will be strongly gravitationally red-shifted.   For a black hole with spin parameter  $a/r_g = 0.99$ (these quantities relate to the physical mass $M$ and angular momentum $J$ of the black hole by $r_g = GM/c^2$ and $a = J/Mc$)  the red-shift factor for favourable (near face-on) orientations is $f_\gamma \approx 0.16$ at the ISCO. This means that for favourable orientations the large gravitational redshift counteracts the intrinsic disc temperature rise,  significantly reducing the observed peak energy.  A black hole with large mass and near maximal spin is the only combination of parameters that can reduce the observed peak disc energies to the UV scale, producing rapidly evolving UV light curves in the disc dominated phase, while still being able to tidally disrupt a typical low mass star.  The question then is whether this combination of parameters can reproduce the light curves of ASASSN-15lh in quantitative detail.  

\section{Modelling ASASSN-15lh light curves}
We follow the methods developed in MB20, to which the reader should refer for detailed information.  The one difference between the current work and MB20 is that, rather than introducing the disc material as a delta-function ring at a single time $t_D$, here we gradually feed matter into a ring using a source term ${\cal S}_M$ in the governing disc equation (Balbus 2017) 
\beq\label{gov}
{\partial \zeta \over \partial t} = {\mathcal W} {\partial \over \partial r} \left( {U^0 \over U_\phi '} {\partial \zeta \over \partial r}\right) + {\cal S}_M ,
\eeq
{where $\zeta \equiv r \Sigma W^r_\phi /U^0$ allows the evolution equation for the disc surface density $\Sigma(r,t)$, under the influence of a turbulent stress tensor $W^r_\phi$, to be written in a compact form.  We have defined ${\mathcal W} = (1/U^0)^2\left[W^r_\phi + \Sigma\, \partial W^r_\phi/\partial\Sigma\right]$.  The angular momentum gradient is denoted $U_\phi'$, and $U^0$ is the temporal component of the discs four velocity $U^\mu$.  In common with all analytic models of TDE evolution, our model has necessarily been simplified (see section 4.9 of MB20).  If the disc bolometric luminosity remains sub-Eddington, a thin disc model should provide a reasonably accurate description of the disc evolution.}   We use the following prescription for the mass source term:
\beq\label{source}
{\cal S}_M \propto \delta\left(r - 6r_g\right) \, \exp\left[ - \left( { t  - t_1 \over t_2}\right)^2\right] .
\eeq
Here $t_1$ is an effective `disc formation' time, which we expect to coincide roughly with the time of UV re-brightening, and $t_2$ controls the abruptness of disc formation {(Table 1).  The injection process leaves the disc free to relax to larger radii.   This form of the injection function allows the TDE light curves to transition smoothly from the early-time to disc-dominated phases, avoiding discontinuous behaviour that would be present from introducing matter instantaneously. }   Anticipating a large black hole mass,  matter is fed into the innermost disc regions, with a feeding radius $r = 6 r_g$.   (This happens to be the innermost stable circular orbit for Schwarzschild geometry, but this is not relevant here.  We have verified numerically that the choice of $6 r_g$ does not noticeably effect the quality of fit or the best fit parameters.)   The typical radius at which a TDE disc forms is strongly influenced by the tidal radius of the disrupted star, $ r_T/r_g \sim M^{-2/3}$.   Discs around larger mass black holes therefore typically form at smaller radii.   

We simultaneously fit the evolving flux in nine different observing bands, across infrared (i), optical (r,V,g,B), and ultra-violet (U, UVW1, UVM2, UVW2) frequencies.   As in MB20, the evolving disc flux is computed at each band's effective frequency.  
We fix the black hole spin to that of a rapid prograde spin $a/r_g = 0.99$, and orientate the disc face-on.  Solutions for different spins and orientation angles are discussed in section \ref{bhparams}.   The remaining parameters are the black hole mass $M$, the accreted mass $M_{\rm acc}$ (in effect a normalisation on the source term ${\cal S}_M$), the viscous evolution timescale of the disc $t_v$, and the time constants $t_1$ and $t_2$ in the matter source term (eq. \ref{source}).   We fit the disc model with a small but finite ISCO stress, $\gamma = 5$ in the notation of Mummery \& Balbus (2019).   We may rule out the presence of a large ISCO stress, which produces UV light curves that evolve too slowly.   Acceptable fits were also found for a vanishing ISCO stress.  The possible relationship between the ISCO stress and spin are relatively unexplored, especially near in the near maximal prograde spin limit.   But a small stress is to be expected:  the distance between the event horizon, where the stress must vanish, and the ISCO is very small near maximal $a/r_g$. 

\begin{table}
\renewcommand{\arraystretch}{1.5}
\centering
\begin{tabular}{|p{2.3cm}|p{2.6cm}|}
\hline
$t_1$ (days) & $t_{\rm peak}+  92^{+5}_{-15}$ \\ \hline 
$t_2$ (days) & $50^{+15}_{-20}$ \\ \hline 
$t_v$ (days) & $16.3^{+1.7}_{-1.5}$  \\ \hline 
$ M_{\rm acc}/ M_{\odot} $ & $ 7.3^{+0.5}_{-0.3} \times 10^{-2}$ \\ \hline
$M/ M_{\odot}$ & $1.4^{+0.2}_{-0.35} \times 10^{9}$ \\ \hline
$a/r_g$${}^\dagger$ & 0.99 \\ \hline
$\bar\chi^2_{\rm min}{}^*$ & $2.49$ \\ \hline
\end{tabular}
\caption{Best fit model parameters. ${}^\dagger$ The black hole spin was fixed. ${}^*$ The formally large reduced chi-squared is discussed below.}
\label{table1}
\end{table}

\begin{figure}
  \includegraphics[width=.5\textwidth]{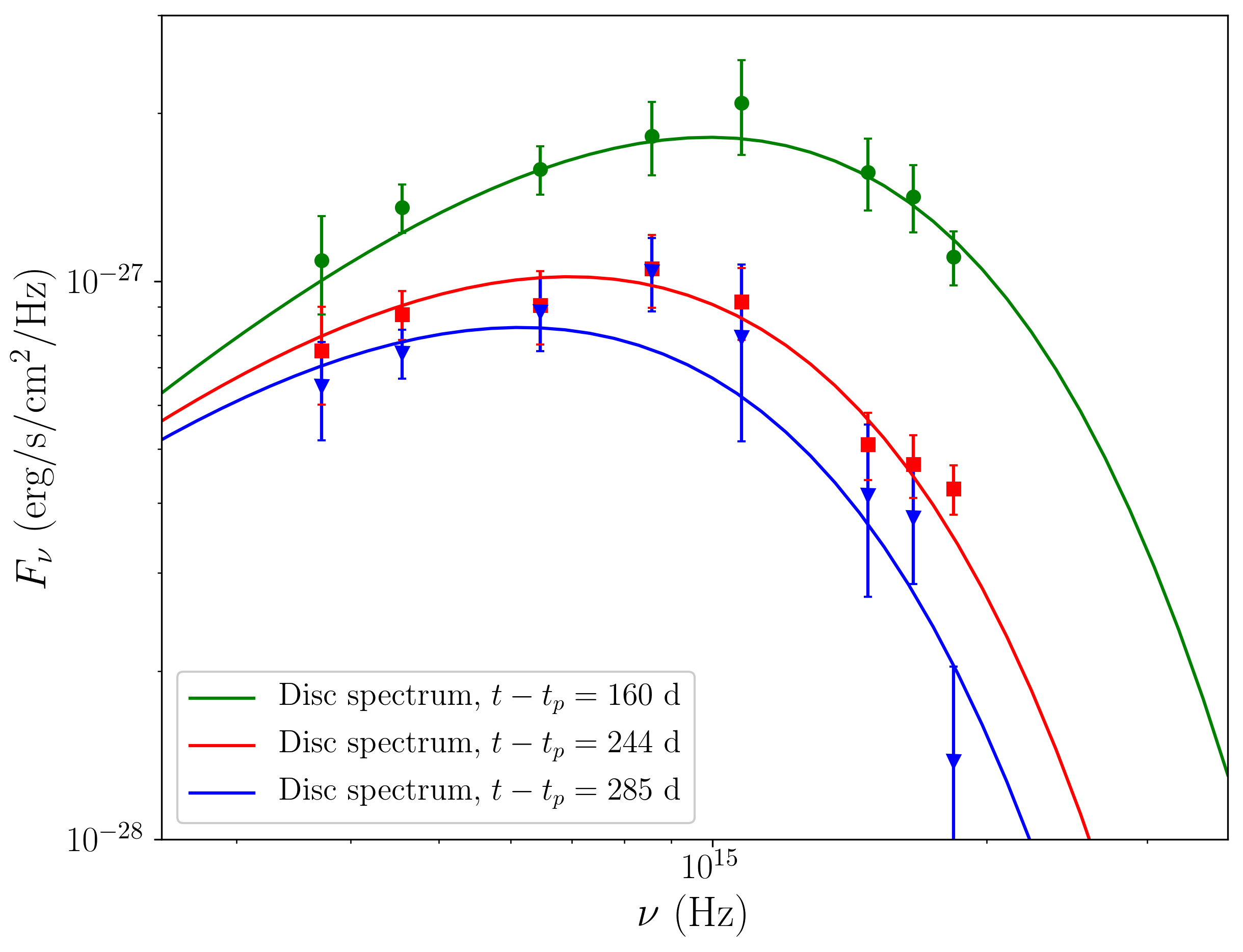} 
 \caption{The best fit disc spectrum and nearly-simultaneous observations at three different times, denoted on plot. }
 \label{spectrum}
\end{figure}

\begin{figure*}
  \includegraphics[width=.76\textwidth]{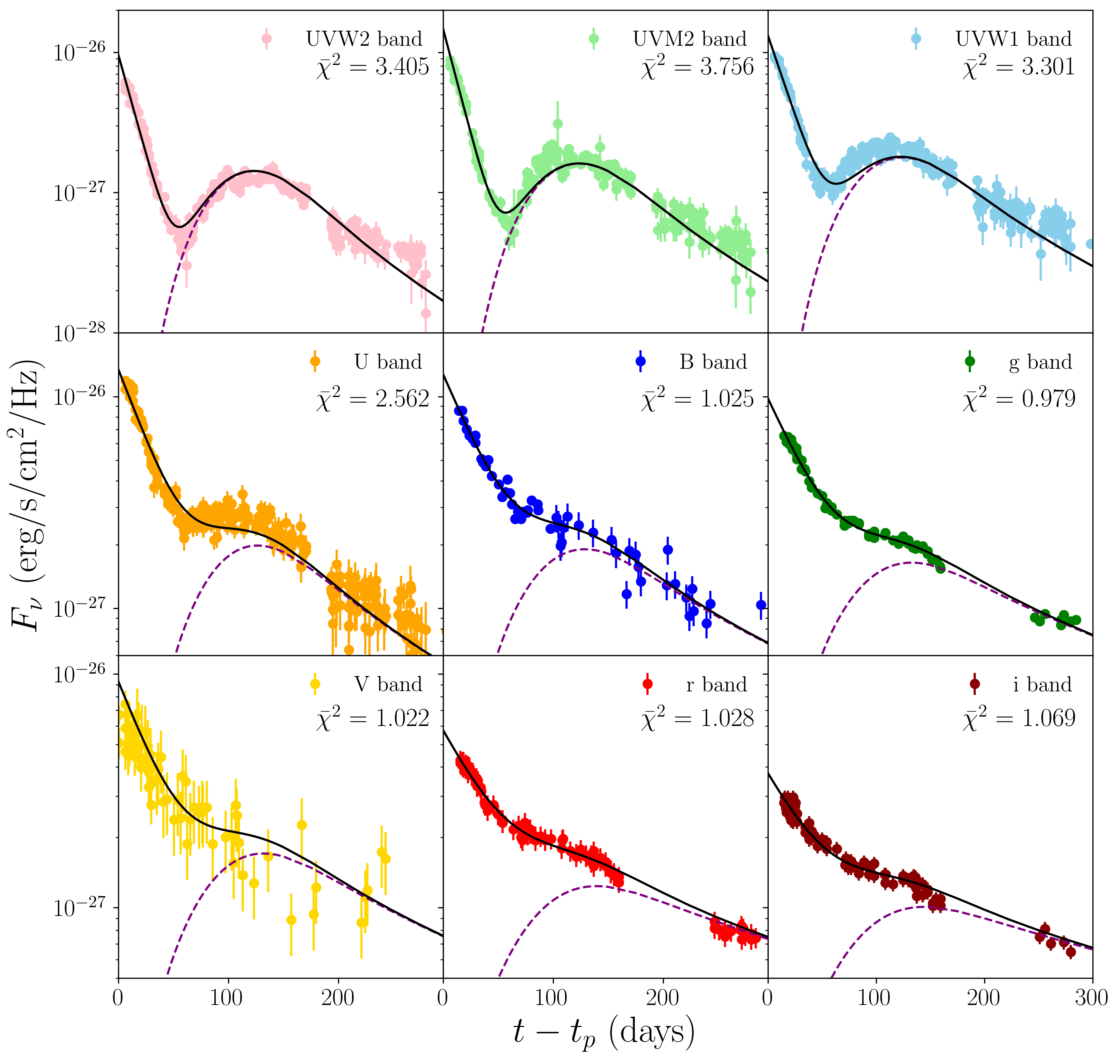} 
 \caption{The disc model and observed ASASSN-15lh light curves. In each figure the contribution from the evolving accretion disc is displayed by a purple dashed curve. The sum of this disc flux and an exponentially declining flux, relevant at early times, is displayed by black solid curves. {While the curves are fit simultaneously, the reduced $\chi^2$ of each band is also displayed; the main source of model-data discrepancy results from correlated short-timescale fluctuations in the four UV light curves.  } }
 \label{fit}
\end{figure*}

{We have minimised the $\chi^2$ statistic of the nine observed light curves, with a best fit reduced value of $\bar\chi^2_{\rm min} = 3414.7/1370 \,=2.49$ (Table \ref{table1}).}  {This formally large reduced $\chi^2$ statistic is a clear consequence of short-timescale fluctuations present in the four well-sampled UV light curves, and is to be expected in any theoretical model using a smooth functional form for the turbulent stress tensor $W^r_\phi$.   This very standard approach implicitly averages over rapid turbulent variations.   Short timescale fluctuations are likely to be highly correlated so that accurately assessing the statistical significance of the fit is not straightforward.   We have therefore used $\chi^2$ minimisation as a sensible guide towards finding a best fit,  but as $\chi^2$ does not have a gaussian normal distribution, we have not attempted a quantitative assessment of fit, leaving the plots to speak for themselves.}    The principal constraint on the black hole mass comes from the simultaneous fitting of the highest and lowest photon energy light curves (MB20):   If the black hole mass is too large, the solution is too bright in the IR;  too low a mass and the solution is too bright in the UVW2 band.  

Figure \ref{spectrum} depicts the disc spectrum at three different times during the disc dominated phase (denoted on plot).  Over-plotted are (nearly) simultaneous observations in a number of different observing bands.   These are well fit by the disc model.   As was argued in section \ref{scaling}, to reproduce the rapidly evolving ultra-violet light curves it is crucial that the observed peak frequency of the disc spectrum lies at $\nu_p \sim \nu_{UV} \sim 10^{15}$ Hz. 

Figure \ref{fit} shows the fitted light curve in the nine different observing bands.  The contribution from the disc at each frequency is shown by a purple dashed line. The disc component is separate from the observed flux at the earliest times: the `total light curve' is found by adding an exponentially decaying early time component to the disc.  The early time behaviour is modelled by:
\beq
F_{\nu}   = A \exp\left(-t/\tau \right),\qquad {(\rm early \, time)}
\eeq
each band with its own $A$ and $\tau$. {The physical origin of the early time UV emission is beyond the scope of our model; it could result, e.g., from outflows present in the early stages of the TDE. }
In figure \ref{fit} the combined disc and exponential flux is shown by a solid black line.  We note that exponential profiles are required to accurately model the rapid early time evolution of the UV bands. We find three different timescales for the early time emission.  The highest photon energy light curves evolve quickest, with $\tau_{\rm w1, m2, w2} \simeq 15$ d, whereas the emission at lower photon energies decay more slowly:  $\tau_{\rm u, b, g, v} \simeq 35$ d, and $\tau_{r, i} \simeq 51$ d. 

After $\sim 100$ days all light curves are dominated by the disc.  The subsequent evolution is then dependent upon the explicit observing frequency, with higher energy light curves evolving more rapidly.  The (highest photon energy) UVW2 light curve decays by an order of magnitude over an interval where the IR light curve decays by a factor of only $\sim 1.4$.  This is in accord with the MB20 analysis.  Clearly, an evolving disc model describes the properties of all nine observed light curves very well.

The bolometric luminosity of the disc solution peaks at $\simeq 3.5 \times 10^{45}$ erg/s, {making this disc sub-Eddington at peak ($ 0.02 L_{\rm Edd}$). }  The time-integrated radiated energy equals $E_{\rm rad} \simeq 6 \times 10^{52}$ erg. At times beyond $t  > t_p +  400$ days, the observed flux in the four UV bands deviates from the disc model, remaining significantly brighter than predicted.  This is presumably due to an additional, more slowly decaying, component.  This may be due to weak AGN activity (Kr{\" u}hler {\it et al.} 2018), additional material from the initial disruption returning to the disc, or even a state transition within the disc at low Eddington ratio.  
\subsection{Black hole properties}\label{bhparams}
The source ASASSN-15lh is located at the centre of a massive red galaxy with limited star formation (L16).  There are four existing estimates of the mass of the central black hole in ASASSN-15lh, which are based upon distinct, well-established galactic scaling relations.  Two estimates are derived from properties of the galactic bulge (L16), while two follow from velocity dispersion measurements (Kr{\" u}hler {\it et al.} 2018, Wevers {\it et al.} 2018). The existing estimates for the BH mass are: $\log_{10}\left(M\right) = 8.88 \pm 0.60, \, (M:M_{\rm bulge}, \, {\rm L}16)$, $\log_{10}\left(M\right) = 8.50 \pm 0.52, (M:L_{\rm bulge},  \, {\rm L}16)$, $\log_{10}\left(M\right) = 8.32 \pm 0.31, (M:\sigma ,  \, {\rm W}18)$ and $\log_{10}\left(M\right) = 8.70 \pm 0.41, \, (M:\sigma ,  \, {\rm K}18)$.
Our best fitting black hole mass $\log_{10}(M)=9.15^{+0.05}_{-0.13}$ (Table 1) is consistent with three of these four estimates of the black holes mass, and is of course obtained by independent means. 
This good agreement strengthens the argument for a TDE origin of the source. 

The implied large black hole mass points to a star with a mass in excess of $1M_\odot$ being disrupted.   The analytical model of Kesden (2012) requires a main sequence star with mass $M_\star \simeq 2.6_{-0.9}^{+0.6} M_{\odot}$ for a full disruption.  (The error range here corresponds to the error range on the fitted black hole mass).  
Other models give somewhat lower stellar masses for a full disruption (Ivanov {\it et al}. 2006, Sponholz 1994),  but the best estimates point to a stellar mass of at least $M_\star \sim 2 M_{\odot}$ for full disruption.   

{We have also fit the ASASSN-15lh light curves for a full range of black hole spins, including an $a=0$ Schwarzschild spacetime, independently of any other considerations.  The best fit central masses associated with smaller spins were systematically lower, but always with $M > 9 \times 10^8 M_\odot$.   In general, smaller black hole spins produced poorer fits to the observed light curves.   A Schwarzschild black hole (best fit mass $M = 9\times10^8M_{\odot}$) produced the worst fit $\bar\chi^2_{\rm min} = 4284.2/1370 = 3.13$, and would require a $28.3M_{\odot}$ solar-type star for disruption (Kesden 2012, fig \ref{spin}).  This makes a low spin value very unlikely.  
For larger spins, the quality of fit is similar to the near-maximal $a/r_g = 0.99$ case.  When $a/r_g = 0.9$, the best fitting light curve has: $\bar \chi^2_{\rm min} =  3383.1/1370 = 2.47$, with a very similar best fit black hole mass $M = 1.35\times10^9 M_{\odot}$.  However even this slight reduction from maximal spin now requires a $8.3M_{\odot}$ star for a disruption, compared to $M_\star = 2.6M_{\odot}$ for $a/r_g = 0.99$.     From the light curves alone we cannot rule out the possibility of an extremely massive star being disrupted by a more moderately rotating black hole.    The strong dependence of TDE rate on stellar mass ($\propto M_\star^{-1.89}$ as noted above) clearly favours a near maximal spin. }

The light curves of \S 3 were produced assuming a face-on inclination. The best fit parameters are, in fact, only weakly sensitive to the disc-observer inclination, with only minor changes to the fitted parameters for $\theta < 20^o$.  For higher inclination angles, larger black hole masses were required.   
Taking the uppermost black hole mass allowed from the galactic bulge range as a hard upper limit, we may constrain the inclination to $\theta < 60^o$. 

Given that a relatively massive ($\sim 2M_{\odot}$) star has been disrupted,  it is somewhat surprising that the disc component of the evolving light curves is powered by only a relatively small accreted mass $M_{\rm acc} \simeq  0.07 M_{\odot}$.     It is important to note that this accreted mass value does not include the matter required to power the (extremely bright) early time emission (observed in all bands), or the late time UV emission.  Rather, the disc mass represents only a lower bound on the debris mass.  Given the likely highly relativistic orbit of the disrupted star, it may well be that a significant portion of the stellar mass was directly accreted in the earliest stages.  It has also been argued (Guillochon \& Ramirez Ruiz 2015) that relativistic orbital precession of the tidal disruption streams (induced by a large black hole spin) can lead to inefficient disc formation.

\subsection{Radio and X-ray observations}
No radio emission has been detected from ASASSN-15lh, with upper limits of order $\nu L_\nu \lesssim 10^{38}$ erg/s (L16,  Kool {\it et al}. 2015).
Given the likely near maximal spin of its central black hole, it is perhaps surprising that a jet is not observed.  (The observed radio emission is at least 2 to 3 orders of magnitude lower than the bright, jetted TDE Swift J1644+57 [Levan {\it et al}. 2011].)    Moreover, unlike other jetted TDEs, ASASSN-15lh is X-ray dim.   Initially undetected at X-ray energies (L16),  X-ray emission has been more recently detected by Margutti {\it et al}. (2017) at the level $L_X \sim 10^{41}$--$10^{42}$ erg/s.    A pure disc model is not able to produce even this level of X-ray flux, as the spectrum peaks in the UV and falls off sharply.    With only a handful of detected photons, it is difficult to infer spectral properties, but the emission appears to be soft (Margutti {\it et al}. 2017).     Kr{\" u}hler {\it et al.} (2018) note that spectral features of the ASASSN-15lh host galaxy indicate the presence of a weak AGN, which they argue might be the source of the observed X-rays, but at present the origin of the X-rays is poorly understood.    We note here that the majority of TDEs do not appear to be associated with jets.   It is not yet clear whether the absence of jet emission is an intrinsic or a misalignment effect.
\section{Conclusions}     
We have modelled the extremely luminous transient ASASSN-15lh at nine separate frequencies (simultaneously), and have shown that a relativistic thin disc model reproduces the data well.  {Our methodology has been very simple.   Two pathways to improving the model accuracy, for example, are the treatment of atmospheric radiative transfer physics {(e.g, Done {\it et al.} 2012)}; and ii) the inclusion slim disc dynamics (Abramowicz {\it et al}.\ 1988).  }
The best fit is obtained for a central hole mass of $\simeq 10^9M_\odot$ and a near maximal spin parameter $a/r_g=0.99$, both strikingly large and presumably the result of significant accretion.     The deduced mass is consistent with independent observations based on galactic bulge properties and velocity dispersions.   The overall quality of the fit {is compelling evidence} for a TDE interpretation for ASASSN-15lh.   This would be the first TDE to be observed in the disc dominated state at optical and infrared wavelengths.   
\bigskip

\noindent {\bf ACKNOWLEDGEMENTS}

\smallskip
\noindent This work is partially supported by STFC grant ST/S000488/1, and the Hintze Family Charitable Foundation.   It is a pleasure to acknowledge useful conversations with Nick Stone, and very constructive comments from our referee. 

\label{lastpage}

\end{document}